\documentclass[twocolumn, switch]{article} 

\usepackage{preprint}

\usepackage{amsmath, amsthm, amssymb, amsfonts}

\usepackage[numbers,square]{natbib}
\usepackage[utf8]{inputenc}	
\usepackage[T1]{fontenc}	
\usepackage{xcolor}		
\usepackage[colorlinks = true,
            linkcolor = purple,
            urlcolor  = blue,
            citecolor = cyan,
            anchorcolor = black]{hyperref}	
\usepackage{booktabs} 		
\usepackage{nicefrac}		
\usepackage{microtype}		
\usepackage{lineno}		
\usepackage{pifont}
\usepackage{float}			

\usepackage{lipsum}		

\usepackage{newfloat}
\DeclareFloatingEnvironment[name={Supplementary Figure}]{suppfigure}
\usepackage{sidecap}
\sidecaptionvpos{figure}{c}

\usepackage{titlesec}
\titlespacing\section{0pt}{12pt plus 3pt minus 3pt}{1pt plus 1pt minus 1pt}
\titlespacing\subsection{0pt}{10pt plus 3pt minus 3pt}{1pt plus 1pt minus 1pt}
\titlespacing\subsubsection{0pt}{8pt plus 3pt minus 3pt}{1pt plus 1pt minus 1pt}

\usepackage{tikz,xcolor,hyperref}

\definecolor{lime}{HTML}{A6CE39}
\DeclareRobustCommand{\orcidicon}{
	\begin{tikzpicture}
	\draw[lime, fill=lime] (0,0)
	circle [radius=0.16]
	node[white] {{\fontfamily{qag}\selectfont \tiny ID}};
	\draw[white, fill=white] (-0.0625,0.095)
	circle [radius=0.007];
	\end{tikzpicture}
	\hspace{-2mm}
}
\foreach \x in {A, ..., Z}{\expandafter\xdef\csname orcid\x\endcsname{\noexpand\href{https://orcid.org/\csname orcidauthor\x\endcsname}
			{\noexpand\orcidicon}}
}

\title{Incremental Federated Learning for Intrusion Detection in IoT Networks under Evolving Threat Landscape}

\usepackage{xwatermark}
\newwatermark[firstpage,color=gray!60,angle=90,scale=0.32, xpos=-4.05in,ypos=0]{\href{https://doi.org/}{\color{gray}{Publication doi}}}
\newwatermark[firstpage,color=gray!60,angle=90,scale=0.32, xpos=3.9in,ypos=0]{\href{https://doi.org/}{\color{gray}{Preprint doi}}}
\newwatermark[firstpage,color=gray!90,angle=0,scale=0.28, xpos=0in,ypos=-5in]{*correspondence: \texttt{muaan.ur@taltech.ee}}

\usepackage{authblk}

\author[1\thanks{\tt{asmith@college.edu}}]{Muaan Ur Rehman}
\author[2]{Hayretdin Bahsi\orcidB{}}
\author[1]{Rajesh Kalakoti\orcidC{}}

\affil[1]{Centre for Digital Forensics and Cyber Security, Department of Software Science, Tallinn University of Technology, Tallinn, Estonia}
\affil[2]{School of Informatics, Computing, and Cyber Systems, Northern Arizona University, United States}

\begin{document}

\twocolumn[ 
  \begin{@twocolumnfalse} 

\maketitle

\begin{abstract}
The expansion of Internet of Things (IoT) devices has increased the attack surface of networks, necessitating a robust and adaptive intrusion detection systems. Machine learning based systems have been considered promising in enhancing the detection performance. Federated learning settings enabled us to train models from network intrusion data collected from clients in a privacy preserving manner. However, the effectiveness of these systems can degrade over time due to concept drift, where patterns in data evolve as attackers develop new techniques. Realistic detection models should be non-stationary, so they can be continuously updated with new intrusion data while maintaining their detection capability for older data. As IoT environments are resource constrained, updates should consume minimal computational resources. This study provides a comprehensive performance analysis of incremental federated learning in enhancing the long term performance of non stationary IDS models in IoT networks. 
Specifically, we propose LSTM models within a federated learning setting to evaluate incremental learning approaches that utilize data and model-based measures against catastrophic learning under drift conditions. Using the CICIoMT2024 dataset, which includes various attack variants across five major categories, we conduct both binary and multiclass classification to provide a granular analysis of the intrusion detection task. Our results show that cumulative incremental learning and representative learning provide the most stable performance under drift, while retention-based methods offer a strong accuracy and latency trade off. The study offers new insights into the interplay between training strategy performance and latency in dynamic IoT environments, aiming to inform the development of more resilient IDS solutions considering the resource constraints in IoT devices. 
\end{abstract}
\vspace{0.35cm}

  \end{@twocolumnfalse} 
] 



\section{Introduction}
\label{sec:introduction}

The rapid growth of the IoT has led to an increase in connected devices, making it a vital part of daily life while simultaneously heightening security risks. IoT security is critical, given the potential for financial damage or threats to human safety \cite{lopez2023machine}.

Machine learning and deep learning are widely incorporated into cybersecurity for tasks such as identifying malware, detecting intrusions, detecting online abuse, and recognizing malicious events in centralized settings~\cite{kalakotiTransformer2024}~\cite{kalakoti2025evaluating}. However, ML models in real-world applications can encounter "concept drift", which means the patterns they learned from data change over time~\cite{lu2018learning}. This can occur because the underlying data evolves or the initial data used for training were unstable or poorly defined. 
Furthermore, as cyber threats continuously evolve and new attack vectors emerge in cyberspace due to the continuous attack-defense game between attackers and defenders, it becomes essential to employ non-stationary IDS models that can be regularly retrained or updated to maintain robust detection performance over time. However, IoT devices have limited computational resources, making it essential to select the best training approach. Incremental learning approaches provide a resource-efficient alternative as the previous model is updated with new instances without full retraining. Recent studies, i.e., \cite{JIN202457} \cite{s25144309}, demonstrate that it is possible to have high-performing solutions that combine federated and incremental learning, and they can even outperform traditional centralized intrusion-detection approaches in non-stationary IoT settings.

While FL preserves privacy and reduces the communication load of sensitive information, its performance may significantly drop under data drift. In real-world intrusion detection deployments, new attacks continually emerge, altering the underlying network data distribution. Considering FL frameworks aim to induce models out of devices distributed among various organizations or fields, they may encounter more drift in some clients. Traditional FL training collapses under such drifts, leading to poor generalization. Thus, incremental learning can be an efficient approach in FL settings for updating the model under drifting conditions. However, these methods can suffer from catastrophic forgetting, leading the models to unlearn some concepts as they adapt to changes in the data distribution. It is important to remember old concepts with limited training resources, as networks can be vulnerable to both old and new attacks. Although FL has been explored for intrusion detection~\cite{kalakoti2025federated}, most existing studies develop stationary models that ignore the temporal evolution of attacks, thus failing to address data drift.

In this work, we present a systematic evaluation of a non-stationary federated learning model maintained by incremental learning for intrusion detection tasks in IoT networks. As IoT environments are resource-constrained, we benchmarked the detection and resource-consumption performance of various methods used to prevent catastrophic forgetting.
We used the recent and well-developed CICIoMT2024 IoT dataset~\cite{DADKHAH2024101351} for model training and testing, focusing on previously identified and labeled attacks, rather than zero-day attack detection.

Since intrusion detection data exhibit strong temporal dependencies and evolving traffic patterns, we adopt LSTM as the classifier in this work due to its effectiveness in modeling sequential behaviors and its compatibility with gradient based federated learning.

We construct a realistic incremental learning timeline ($t_{0}$--$t_{6}$) in which new attacks are introduced at each stage, emulating temporal drift conditions in a distributed setting. 
Across this timeline, we benchmark incremental federated learning strategies with varying countermeasures against catastrophic forgetting within the federated learning settings. In our benchmarking, we selected baselines such as a full retrained model, a static model without any updates, and simple incremental learning without any prevention for catastrophic forgetting. We applied varying relearning strategies that select old concepts from broader attack categories or a specific attack category to understand the impact of representation scope. We also included an incremental learning approach that takes the averages of recent model parameters rather than data instances in the new model. All methods are evaluated under both binary and 6-class classification settings using a standardized federated LSTM architecture. 

To the best of our knowledge, this is the first systematic benchmark of incremental federated learning strategies under an explicitly modelled temporal drift scenario in a distributed IoT environment. 
Our results show that representative and retention based incremental FL methods maintain significantly higher performance across evolving attack time periods compared to simple incremental baselines. 
The main contributions of our work are as follows:
\begin{itemize}
    \item We propose a timeline based incremental federated learning framework to systematically analyze the impact of sequential concept drift on intrusion detection systems deployed in distributed IoT environments, considering both binary and multi-class detection settings.
    \item We conduct an extensive empirical evaluation of multiple incremental federated learning strategies using an LSTM based IDS on the CICIoMT2024 dataset, highlighting accuracy vs latency trade offs.
\end{itemize}

The rest of the paper is organized as follows. Section \ref{sec:related_work} reviews the related work and gives background information about the impact of incremental federated learning under evolving threats. Section \ref{sec:DsAndExpSetup} describes the dataset and experimental setting. Section \ref{sec:methodology} presents the methodology used in our paper. In Section \ref{sec:results_discussions} , we show our results and discuss them. Finally, Section~\ref{sec:conclusion} concludes the paper and discusses future directions.

\section{Related Work}\label{sec:related_work}

FL has recently gained significant attention for its effectiveness in IoT applications. Recent studies have explored the application of FL to enhance ID\cite{hei2020trusted},~\cite{kalakoti2024improving}. A paper~\cite{hei2020trusted} conducted a comprehensive evaluation of various machine learning models, including Random Forest (RF), Decision Trees (DT), Support Vector Machines (SVM), and Multilayer Perceptron (MLP), within federated settings that utilize blockchain for the aggregation process. This study incorporated local training using data from IoT devices alongside the KDDCup99 dataset~\cite{tavallaee2009detailed}. Authors from \cite{mothukuri2021federated} introduces a decentralized  FL approach for intrusion detection in IoT networks, using Gated Recurrent units (GRU) as local models and an ensemble of random forest decision trees. This ensemble optimizes the global model’s parameters by aggregating updates from various sources to improve traffic data classification. Based on GRU, a paper~\cite{nguyen2019diot} proposes the creation of communication profiles for IoT devices to detect potential attacks. The data set is generated from real devices and involves Mirai botnet traffic.~\cite{li2021fleam} did not provide performance details, such as the number of participants and the training rounds. ~\cite{khoa2020collaborative} proposes using deep belief networks (DBN) in IoT gateways to detect potential attacks on a specific IoT subnet and then combines the models through FL.  \cite{kalakoti2024explainable} focused on integrating explainability in FL for IDS problems. Specifically, the authors worked on botnet detection for both binary and multiclass classification in industrial IoT settings, using the Fedavg algorithm with Long Short-Term~\cite{kalakoti2025federated} Memory (LSTM) and Deep Neural Network (DNN)~\cite{kalakoti2024explainable} models applied to the n-BaIoT dataset. Furthermore, same authors employed a DNN model across various networks of IDS datasets using the FedAvg algorithm, emphasizing the explainability of federated learning with synthetic data~\cite{kalakoti2025synthetic}.
Authors also explored adapting federated intrusion detection to evolving IoT environments, such as FD-IDS \cite{s25144309}, which applies knowledge distillation to mitigate non-IID client distributions, and FL-IIDS \cite{JIN202457}, which introduces an incremental FL-based IDS for updating models over time.

Although the above studies\cite{hei2020trusted,mothukuri2021federated,tavallaee2009detailed,khoa2020collaborative,nguyen2019diot,kalakoti2025federated,kalakoti2024explainable,kalakoti2025synthetic} demonstrate the potential of FL-based IDS  in IoT systems, they largely assume a static attack distribution and do not tackle the challenges posed by the dynamic, rapidly evolving nature of IoT threat behaviours that highlighting the need for incremental FL approaches capable of continuous learning.

In this work, we focus on Incremental Federated Learning (IFL) for modeling IoT threat behavior under dynamic and evolving attack conditions, enabling the IDS to continuously adapt to newly emerging threats without retraining from scratch. Though FL has been widely explored for IDS, very few studies have investigated the IL paradigm within this domain. 
In a paper~\cite{mahdavi2022itl}, the authors propose a new framework for incremental IDSs that learns without prior knowledge using incremental clustering algorithms and incorporates transfer learning (TL) methods, so that at each model iteration, TL provides incremental knowledge to the target environment, thereby accomplishing incremental learning ~\cite{mahdavi2022itl}. 
An incremental learning model called T-DFNN, which consists of multiple deep feedforward neural network models connected in a tree-like structure, is proposed in \cite{data2021t}. The F1 macro mean of this model at CICIDS2017~\cite{sharafaldin2018toward} achieved more than 0.85 in each training phase. \cite{amalapuram2022continual} evaluates the advantages and disadvantages of two incremental learning methods: Elastic Weight Consolidation (EWC) and Gradient Episodic Memory (GEM). It specifically focuses on scenarios where the data is class-imbalanced, considering both class incremental tasks and domain incremental tasks.

Current FL-based IDS research focuses on improving detection accuracy but largely overlooks the need for continuous adaptation to evolving IoT traffic. Likewise, existing incremental IDS methods are mostly centralized and depend on model restructuring, making them unsuitable for distributed IoT threat environments.

\section{Experimental Setup}\label{sec:DsAndExpSetup}

%

\begin{table}[t]
\centering
\caption{Attack presence and retention across training timeline ($t_0$–$t_5$).  \ding{51} indicates inclusion of an attack class in a time period ;A \ding{51}$^{\downarrow}$ marks the first introduction of an attack class; \ding{51}$^{R}$ marks retained samples from $t_2$ onward ($R\!\in\!\{100,500,1000\}$) in incremental learning by retention.}
\label{tab:train_inclusion_retention}
\renewcommand{\arraystretch}{1.05}
\setlength{\tabcolsep}{3pt}
\scriptsize
\begin{tabular}{lcccccc}
\toprule
\textbf{Sub-attack} & $t_0$ & $t_1$ & $t_2$ & $t_3$ & $t_4$ & $t_5$ \\
\midrule
\textbf{Benign} \\
Benign                              & \ding{51}$^{\downarrow}$ &\ding{51}  & \ding{51}$^{R}$ & \ding{51}$^{R}$ & \ding{51}$^{R}$ & \ding{51}$^{R}$ \\
\midrule
\textbf{MQTT} \\
MQTT\text{-}Malformed\_Data         &                    & \ding{51}$^{\downarrow}$ & \ding{51}$^{R}$ & \ding{51}$^{R}$ & \ding{51}$^{R}$ & \ding{51}$^{R}$ \\
MQTT\text{-}DoS\_Connect\_Flood     &                    & \ding{51}$^{\downarrow}$ & \ding{51}$^{R}$ & \ding{51}$^{R}$ & \ding{51}$^{R}$ & \ding{51}$^{R}$ \\
MQTT\text{-}DDoS\_Publish\_Flood    &                    & \ding{51}$^{\downarrow}$ & \ding{51}$^{R}$ & \ding{51}$^{R}$ & \ding{51}$^{R}$ & \ding{51}$^{R}$ \\
MQTT\text{-}DDoS\_Connect\_Flood    & \ding{51}$^{\downarrow}$ & \ding{51}  & \ding{51}$^{R}$ & \ding{51}$^{R}$ & \ding{51}$^{R}$ & \ding{51}$^{R}$ \\
\midrule
\textbf{DoS} \\
TCP\_IP\text{-}DoS\text{-}TCP       &                    &     & \ding{51}$^{\downarrow}$ & \ding{51}$^{R}$ & \ding{51}$^{R}$ & \ding{51}$^{R}$ \\
TCP\_IP\text{-}DoS\text{-}ICMP      &                    &     & \ding{51}$^{\downarrow}$ & \ding{51}$^{R}$ & \ding{51}$^{R}$ & \ding{51}$^{R}$ \\
TCP\_IP\text{-}DoS\text{-}SYN       &                    &     & \ding{51}$^{\downarrow}$ & \ding{51}$^{R}$ & \ding{51}$^{R}$ & \ding{51}$^{R}$ \\
TCP\_IP\text{-}DoS\text{-}UDP       & \ding{51}$^{\downarrow}$ &     & \ding{51}& \ding{51}$^{R}$ & \ding{51}$^{R}$ & \ding{51}$^{R}$ \\
\midrule
\textbf{DDoS} \\
TCP\_IP\text{-}DDoS\text{-}SYN      &                    &     &             & \ding{51}$^{\downarrow}$ & \ding{51}$^{R}$ & \ding{51}$^{R}$ \\
TCP\_IP\text{-}DDoS\text{-}ICMP     &                    &     &             & \ding{51}$^{\downarrow}$ & \ding{51}$^{R}$ & \ding{51}$^{R}$ \\
TCP\_IP\text{-}DDoS\text{-}UDP      & \ding{51}$^{\downarrow}$ &     &             & \ding{51} & \ding{51}$^{R}$ & \ding{51}$^{R}$ \\
TCP\_IP\text{-}DDoS\text{-}TCP      &                    &     &             & \ding{51}$^{\downarrow}$ & \ding{51}$^{R}$ & \ding{51}$^{R}$ \\
\midrule
\textbf{Recon} \\
Recon\text{-}Ping\_Sweep            &                    &     &             &             & \ding{51}$^{\downarrow}$ & \ding{51}$^{R}$ \\
Recon\text{-}VulScan                &                    &     &             &             & \ding{51}$^{\downarrow}$ & \ding{51}$^{R}$ \\
Recon\text{-}OS\_Scan               &                    &     &             &             & \ding{51}$^{\downarrow}$ & \ding{51}$^{R}$ \\
Recon\text{-}Port\_Scan             & \ding{51}$^{\downarrow}$ &     &             &             & \ding{51}& \ding{51}$^{R}$ \\
\midrule
\textbf{Spoofing} \\
ARP\_Spoofing                       & \ding{51}$^{\downarrow}$ &     &             &             &             & \ding{51}\\
\bottomrule
\end{tabular}
\end{table}

This section describes the dataset \cite{DADKHAH2024101351}, the construction of the temporal evaluation timeline, and the experimental settings used to study concept drift in federated learning–based intrusion detection systems. We detail how attack families are organized into consecutive time periods, the data partitioning across federated clients, and the configuration used for both binary and multi-class intrusion detection experiments.

\subsection{Dataset}

We conduct the incremental federated learning based drift analysis to the CICIoMT2024 dataset \cite{DADKHAH2024101351}, which focuses on Internet of Medical Things (IoMT) devices operating in healthcare settings. This dataset provides a comprehensive benchmark containing multiple attack variants, designed to facilitate intrusion detection and enhance IoMT cybersecurity. It includes network traffic from 40 devices (25 real and 15 simulated) communicating over various protocols such as Wi-Fi, MQTT, and Bluetooth. In total, 18 cyberattacks were simulated and organized into five main categories i.e. MQTT, DoS,DDoS, Reconnaissance,and Spoofing.
The DDoS and DoS categories include traditional flooding attacks (e.g., SYN, TCP, UDP, and ICMP floods) designed to exhaust device or network resources. Reconnaissance attacks, such as port scans, OS scans, ping sweeps, and vulnerability scans, emulate adversarial information gathering phases. MQTT  attacks involve connect floods, publish floods, and malformed MQTT packets aimed at disrupting broker communication. Spoofing attacks, including ARP spoofing, which exploit network-layer weaknesses to manipulate or intercept legitimate traffic.
The dataset contains 45 features in total, encompassing header and flow metadata, TCP/IP flag indicators, protocol identifiers, statistical descriptors, and additional network-level attributes relevant to intrusion detection

The raw dataset was first cleaned to remove missing or NA rows. The attack \texttt{MQTT-DoS-Publish\_Flood} was completely removed from all experiments, as no valid instances were found in the cleaned dataset after preprocessing. The dataset was stratified using the \texttt{Attack} column to preserve proportional representation of attacks. The data were then divided into 80\% training and 20\% testing subsets using stratified sampling.  
Min–Max normalization was applied for feature scaling, and categorical attributes were label-encoded to ensure consistent mappings across all time periods and clients.


\subsection{Timeline Construction}

To simulate different drift conditions, both training and testing datasets were divided into six or seven temporal time periods.

In the binary setting, we start directly from \(t_1\), which contains two well-formed classes (Benign vs. MQTT), enabling meaningful supervised learning from the very first stage. In contrast, the 6-class classification task requires full coverage of all six categories (Benign + five attack families). Since \(t_1 \),   does not include every attack family (except representative learning), we introduce \(t_0 \),   as a controlled baseline containing one representative sub-attack per category. This guarantees that the model begins with a complete label space, ensuring fair and stable 6-class learning before drift is introduced at later time periods.

Starting from \(t_1\), new attack families were introduced at each subsequent time period to emulate evolving threat distributions. Specifically, the MQTT family was introduced at \(t_{1}\), DoS at \(t_{2}\), DDoS at \(t_{3}\), Recon at \(t_{4}\), and Spoofing at \(t_{5}\).  The final time period (\(t_5\)) contains all attack categories, representing the complete attack space after full evolution. Time period \(t_{6}\) was used only for testing (no training) and contained the same attack families as \(t_{5}\) to evaluate model generalization after full exposure. Our attack distribution is totally different for every training strategy. We added the attacks from the categories according to the dataset attack class imbalance behaviour. We added the majority attack classes in the earlier time periods because their large sample sizes provide stable learning signals, whereas introducing minory attacks classes early would produce unstable models, making the drift urealistic and unreprestnative. (Table~\ref{tab:train_inclusion_retention}) shows how we simulate drift condition for cumulative incremental learning and incremental learning by retention. \ding{51} in the table indicates inclusion of an attack class in a time period ;A \ding{51}$^{\downarrow}$ marks the first introduction of an attack class; \ding{51}$^{R}$ marks retained samples from $t_2$ onward ($R\!\in\!\{100,500,1000\}$) in incremental learning by retention.
\subsection{Experimental Settings}

We implemented the FL setting using the Flower framework. Data were distributed across five clients using IID partitioning strategies to isolate and analyse the effects of temporal concept drift independently from data heterogeneity. This controlled setup ensures that observed performance degradation is attributable to evolving attack distributions rather than client-level data imbalance

To ensure a fair and temporally consistent construction of the incremental timeline, all attack classes were first divided chronologically into equal-sized temporal segments. For a setup with six time periods, each attack type was partitioned into six disjoint subsets, e.g. an attack class with 462,480 instances (e.g., \texttt{DoS\_TCP}) contributes approximately 77080 samples to each time period. Minority classes follow the same procedure i.e. an attack with 6877 samples (e.g., \texttt{MQTT\_Malformed}) provides roughly 1146 samples per time period.

After temporal segmentation, a class-capping mechanism was applied to enforce controlled and balanced federated training. For every time period, each available attack class was limited to at most 10,000 training samples and 2,000 testing samples. This prevents majority classes from dominating the federated optimization process while ensuring that minority classes remain fully represented.

The capped samples at each time period were then distributed across five federated clients using IID sampling. Thus, for every time period and every learning strategy, each client receives an equal proportion subset of the temporally segmented and capped attack classes.
Each client dataset was normalized using min-max scaling, label encoded consistently across all time periods, and split into training and validation data. The whole federating learning split across time periods mainly consists of the training data (60\%), client side testing (10\%) and validation data (10\%) , to validate server rounds. 
Federated aggregation was performed using the standard federated averaging (FedAvg) strategy, where locally trained model parameters are averaged proportionally to client data size at each communication round.
The remaining 20\% test data were reserved for global model testing, split equally into time periods.

This pipeline guarantees that (i) temporal drift is simulated consistently, (ii) no class overwhelms the training distribution, and (iii) all federated clients participate with balanced and comparable datasets.


All experiments used a multi-layer \textbf{LSTM} model with five hidden layers, each containing 128 hidden units. These architectural and training hyperparameters were selected based on preliminary centralised experiments, where this configuration consistently yielded the best performance on CICIoMT2024 dataset.
The model accepted 45 normalized features as input and produced six categorical outputs corresponding to 
\textit{Benign}, \textit{MQTT}, \textit{DoS}, \textit{DDoS}, \textit{Recon}, and \textit{ARP\_Spoofing}.  
Federated training was performed over 15 global communication rounds with five clients, each running 100 local training epochs per round in each time period. This setup represents an upper-bound training scenario executed on an aggregation capable node, ensuring stable convergence across evolving time periods. The learning rate was fixed at 0.001, and the batch size was set to 16.

Model performance was assessed using standard metrics derived from the confusion matrix, including 
Accuracy, Precision, Recall, F1-score, False Alarm Rate (FAR). Trends across Precision, Recall and F1-score were consistent with accuracy results, therefore, we report the overall accuracy ($\text{Acc}$) at each time period in the paper.
The overall accuracy ($\text{Acc}$) is computed using micro-averaging as the ratio of correctly predicted samples to the total number of test samples.

\begin{equation}
\text{Acc} = \frac{\sum_{i=1}^{C} \text{TP}_i}{\sum_{i=1}^{C} (\text{TP}_i + \text{FP}_i + \text{FN}_i)}
\end{equation}
where $\text{TP}_i$, $\text{FP}_i$, and $\text{FN}_i$ denote the true positives, false positives, and false negatives for class $i$, and $C$ represents the total number of classes.

Training and inference latency (in seconds) 
were recorded to evaluate computational efficiency.  
This comprehensive setup enables consistent, reproducible comparison of drift-handling capabilities 
across all federated learning configurations.

\section{Methodology}\label{sec:methodology}

This section presents the proposed federated learning methodology and the incremental learning strategies employed to handle concept drift under evolving attack conditions. Specifically, we describe the federated learning framework and aggregation process, followed by the incremental and continual learning strategies designed to adapt the model across successive time periods.

\subsection{Federated Learning}
Federated Learning (FL) is a decentralized ML approach introduced by Google~\cite{mcmahan2017communication} that enables model development using data from multiple parties while preserving data ownership and privacy. FL has gained popularity and seen significant advancements along with real-world applications.

We now formally describe the federated learning framework and introduce the notation used throughout this work.We denote $N$ as the number of clients, where each client is represented by $c_1, c_2, \ldots, c_N$. The set of samples owned by client $c_i$ is denoted as $U_{c_i}$, the feature space as $X_{c_i}$, the label space as $Y_{c_i}$, and the dataset as $\mathbb{D} _{c_i} = \{ (u_{c_i}^{(j)}, x_{c_i}^{(j)}, y_{c_i}^{(j)}) \}_{j=1}^{|\mathbb{D}_{c_i}|}$. Each data point $(u_{c_i}^{(j)}, x_{c_i}^{(j)}, y_{c_i}^{(j)})$ indicates that client $c_i$ owns the sample $u_{c_i}^{(j)}$, with features $x_{c_i}^{(j)}$ and label $y_{c_i}^{(j)}$. In a network traffic ID scenario, $U$ represents individual network flows, $X$ includes features like packet size and protocol type, and $Y$ indicates if traffic is benign or malicious. Based on how the data \((X,Y, U)\) is partitioned across clients, FL can be categorized into three types by ~\cite{yang2019cdfederated}.  

The client–server architecture is the most popular HFL architecture, and the Federated Averaging (FedAvg) algorithm~\cite{mcmahan2017communication} is based on it. The optimization objective of HFL is as follows:

\begin{equation}
\min_{\theta} L(\theta) = \sum_{i=1}^{N} L_{c_i}(\theta) 
= \sum_{i=1}^{N} \frac{1}{|D_{c_i}|} \sum_{j=1}^{|D_{c_i}|} 
\ell\!\left(x^{(j)}_{c_i},\, y^{(j)}_{c_i};\, \theta\right)
\label{eq:hfl_obj}
\end{equation}

where $\theta$ represents the model parameters, and $L(\theta)$ is the global optimisation objective.  
The term  $ L_{c_i}(\theta) = \frac{1}{|D_{c_i}|} \sum_{j=1}^{|D_{c_i}|} 
\ell\!\left(x^{(j)}_{c_i},\, y^{(j)}_{c_i};\, \theta\right) $ is the optimization objective of client $c_i$ based on its local dataset $D_{c_i}$, with $\ell(x, y; \theta)$ denoting the loss function, such as cross-entropy or mean-squared error. In this work, we have used FedAvg for our experimental settings. 
I



\subsection{Incremental Learning and Drift Scenarios}

A total of six  training strategies were implemented to evaluate drift adaptability in federated learning environments. At each time period \(t_i\) the model update begins from the last check point obtained from the previous time period \(t_{i-1}\), ensuring incremental learning.
In all 6-class classification methods, \(t_0\) serves as the baseline time period, since both training and testing are performed on the same attack set in this period.  
The key difference across methods arises from how attack classes are included, excluded, or limited at each time period.  

For all incremental learning in our setup, the evaluation follows a cross-time period approach, i.e. 
the model trained on \(train_{t_0}\) is tested on \(test_{t_0}\) (no drift) and then on \(test_{t_1}\) (first drift).  
Subsequently, the model trained on \(train_{t_1}\) is tested on \(test_{t_2}\),  
then \(train_{t_2}\) on \(test_{t_3}\), and so on up to \(train_{t_5}\) on \(test_{t_6}\).  
This allows direct measurement of how well each method handles drift when trained on one attack set 
and evaluated on a distinct distribution in the next time period.

\textbf{Static Training:} The model is trained once on the training data at the starting time period, i.e at \(t_1\) (MQTT and Benign) in binary, at \(t_0\) (baseline)  in 6-class classification and evaluated across all subsequent test sets (\(t_0, t_1, \ldots, t_6\)) without retraining. This provides the baseline for drift impact analysis.

\textbf{Cumulative incremental learning:}
The model is trained incrementally at each \(t_i\), incorporating all previously observed attack labels (see Table~\ref{tab:train_inclusion_retention}), with new data points used in each increment. 

In the binary classification setting, the model is first trained at \(t_1\) (Benign and MQTT) and evaluated on the corresponding test split (no drift), followed by evaluation on the next time period \(t_2\) (Benign, MQTT, DoS). At \(t_2\), the model is retrained on the available classes (Benign, MQTT, DoS) and then tested on \(t_3\) (Benign, MQTT, DoS, DDoS). The model trained at \(t_3\) is evaluated on \(t_4\) (Benign, MQTT, DoS, DDoS, Recon), and the model trained at \(t_4\) is evaluated on the full test set at \(t_5\) (Benign, MQTT, DoS, DDoS, Recon, Spoofing). Finally, the model trained at \(t_5\) is tested on \(t_6\), where the same set of six classes is present.

For the 6-class classification, we include a baseline \(t_0\), which provides one representative instance from each of the six classes used in the task, i.e., Benign, MQTT-DDoS-Connect\_Flood, TCP\_IP-DoS-UDP, TCP\_IP-DDoS-UDP, Recon-Port\_Scan, and ARP\_Spoofing. The model trained at \(t_0\) is evaluated on the \(t_0\) baseline and then on \(t_1\). Similar to the binary case, only Benign and MQTT are introduced at \(t_1\).The model trained at \(t_1\) is subsequently evaluated on \(t_2\) (Benign, MQTT, DoS), and the same incremental procedure continues in alignment with the binary classification setup.

\textbf{Simple Incremental Learning:} 
In this type of strategy, only newly introduced attack families are added to the training set at each time period,  representing a pure online learning setup with no memory of earlier attacks. The model is retrained incrementally at each \(t_i\), incorporating all previously observed attack labels (see Table~\ref{tab:train_inclusion_retention}). For the binary classification setup, the model is initially trained at \(t_1\) using Benign and MQTT, evaluated on its corresponding test split (no drift), and then evaluated on \(t_2\) (Benign, MQTT, DoS), which introduces DoS. At \(t_2\), the model is retrained using only the newly encountered classes (Benign and DoS) and then evaluated on \(t_3\)(Benign, MQTT, DoS, DDoS), where the DDoS new class appears. The model updated at \(t_3\) (Benign and DDoS) is then tested on \(t_4\), which introduces Recon. Similarly, the model trained at \(t_4\) (Benign and Recon) is evaluated on the complete test set at \(t_5\), which includes all six attack families. Finally, the model trained at \(t_5\) (Benign and Spoofing) is evaluated on \(t_6\), where the same six classes are present. As discussed earlier in the 6-class classification, we introduce an additonal baseline \(t_0\), containing one representative attack from each of the six categories. The model trained at \(t_0\) is evaluated on both \(t_0\) and \(t_1\). The model updated at \(t_1\) (Benign and MQTT) is then evaluated on \(t_2\), which introduces DoS. The incremental process continues in the same fashion as in the binary classification scenario discussed above.

     \textbf{Representative Incremental Learning:}  
At each time period, newly appearing attack families are added to the training set, while one representative attack from every \emph{other} category is retained. This ensures that the model always trains on a balanced 6-class label space, preventing class disappearance as drift progresses.
In the 6-class setting, training begins at \(t_0\), which contains one representative attack from each of the six categories (Benign + five attack families). This serves as a stable baseline before any drift occurs. At \(t_1\), the full MQTT family is introduced, and the model is updated using all MQTT attacks together with one representative attack from DoS, DDoS, Recon, and Spoofing. The updated model is evaluated on \(t_2\), where the DoS family first appears. At \(t_2\), all DoS attacks are added to the training set, while representative samples from the remaining families are retained.

This incremental process continues similarly across time periods, i.e. at \(t_3\) the full DDoS family is introduced, at \(t_4\) the full Recon family, and at \(t_5\) the complete Spoofing category. In every case, the training set at timeline \(t_i\) includes (i) all attacks newly introduced from a particular category at \(t_i\), and (ii) one representative attack from all other categories. Finally, the model trained at \(t_5\) is evaluated on \(t_6\), which contains the same six attack categories but is used strictly for testing (no further training).

    \textbf{Incremental Learning by Retention:}  
    In this method (See Table~\ref{tab:train_inclusion_retention}), only a limited number of samples (100, 500, or 1000)  are retained
    from previously introduced attacks instead of using the full data for a new increment(Unlike cumulative incremental learning).  This strategy differs fundamentally from representative incremental learning where we were keeping the 6 classes all the time.  Instead of keeping a fixed representative class from each attack category, the model retains only a small, fixed number of samples (e.g., 100, 500, or 1000) from the data observed in earlier time periods. These retained samples act as a compact memory buffer to mitigate forgetting, but they do not preserve categorical balance or full-family coverage.  
    For example, if the model learned MQTT-DoS-Connect\_Flood and Benign at \(t_1\), only limited samples from MQTT-DoS-Connect\_Flood and \textit{Benign} classes 
    are retained during training at \(t_2\), while the rest of the data corresponds to newly introduced attacks.  
    
\textbf{Averaging Incremental Learning Variants:}  
    These are variants of simple incremental learning where the model at each new time period is initialised  by averaging the parameters of previous models.  
    Equal averaging gives uniform weight to all past models. Sample-weighted averaging accounts for the number of samples used in prior training, and EMA applies a decaying weight controlled by a smoothing factor ($\alpha=0.6$). This allows gradual adaptation to drift while mitigating catastrophic forgetting.

\section{Results}
\label{sec:results_discussions}

This section presents the empirical evaluation of all training strategies described in Section~\ref{sec:methodology}. Each approach was assessed under sequential concept drift across time periods $t_{0}$–$t_{6}$ using the CIC-IoMT2024 dataset~\cite{DADKHAH2024101351}. The experiments benchmarked the proposed incremental learning strategies using an LSTM model in a federated setting. We report overall binary and 6-class classification accuracy, per time periodd degradation or recovery trends, and latency metrics to analyse trade-offs between adaptability and computational efficiency.
In the 6-class classification settings at $t_{0}$,  all methods were trained on identical attack class labels to establish a common baseline; subsequent time periods introduced new attack families (MQTT, DoS, DDoS, Recon, and Spoofing) to emulate real-world distributional shifts. The binary classification starts directly from the first increment ( MQTT and Benign)

\begin{figure}[!ht]
  \centering
     \includegraphics[width=0.9\linewidth]{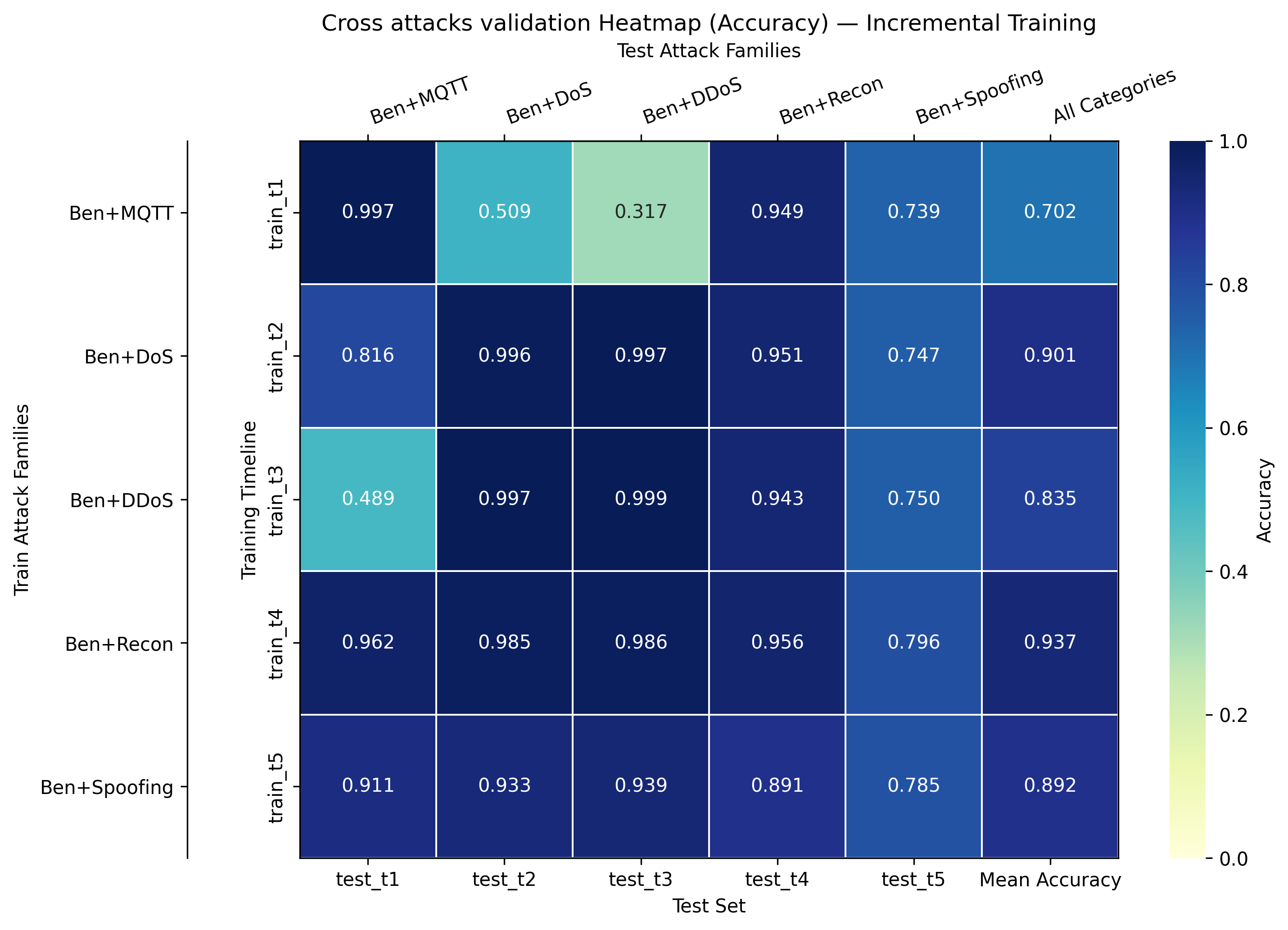}
  \caption{Comparison of algorithms' performance using incremental learning methods over  CICIoMT2024 data set for binary blassification. }
  \label{fig:binary_mean_accruacy}
 \end{figure}

Figure~\ref{fig:binary_mean_accruacy} shows an attack  to attack generalization heatmap produced under our incremental federated learning binary setup.
The rows correspond to the attack families used for training, and the columns correspond to the attack families used for testing.
Each cell reports the classification accuracy when a model trained on one attack distribution is evaluated on another.
This experiment is performed to measure how similar or different the attack families are, i.e., how much concept drift they induce when the model moves from one time period to the next.
The results reveal that MQTT and DDoS exhibit the strongest distributional divergence, as models trained on other attacks perform poorly on these and vice versa.
The last column shows the mean accuracy across all test families, summarising each model’s robustness to unseen attack distributions.

\begin{figure}[!ht]
    \centering
    \includegraphics[width=\linewidth, height=4.4cm]{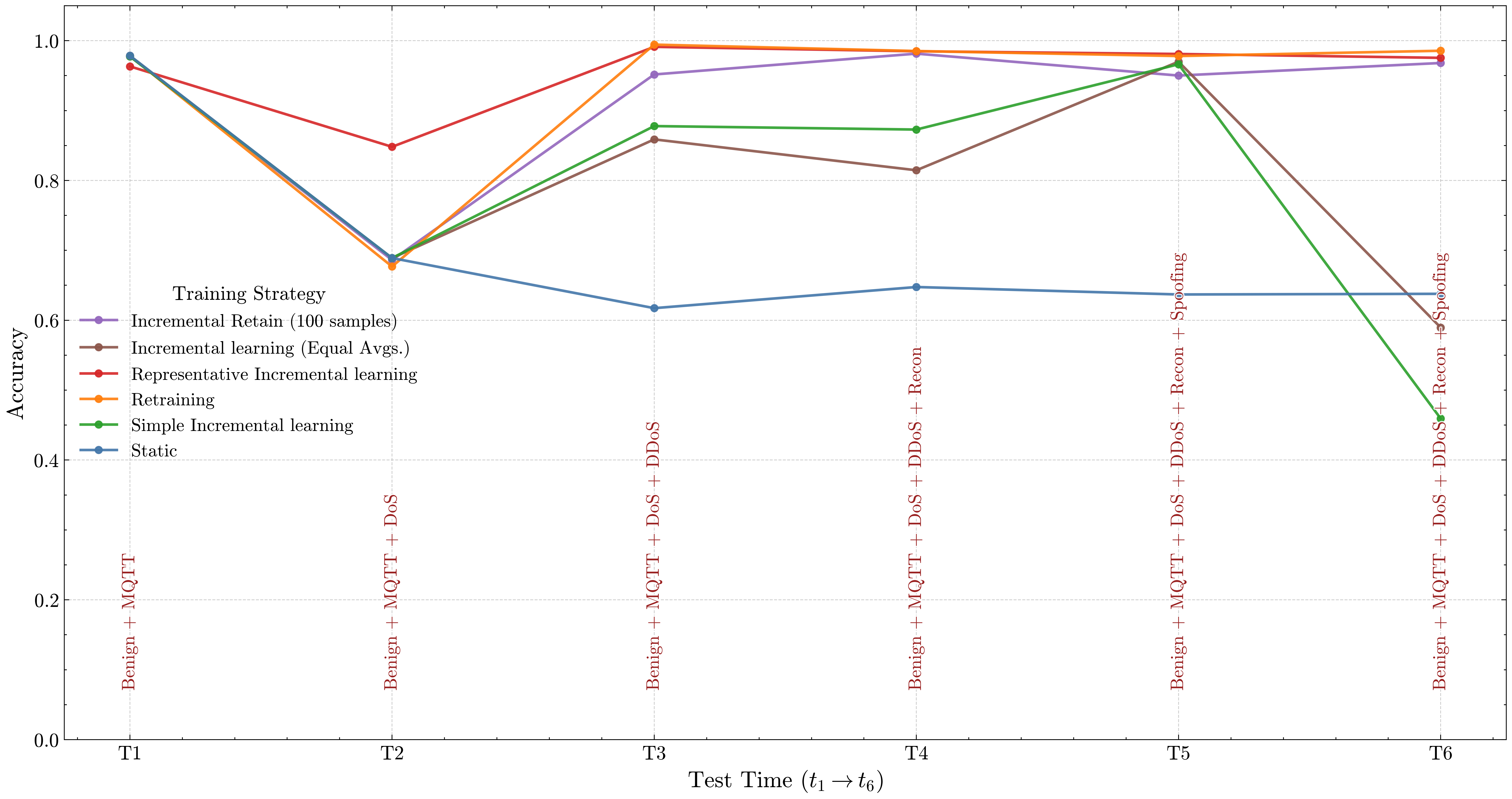}
    \caption{Binary classification across time periods $t_{1}$–$t_{6}$ considering the main six training strategies to benchmark the most drift-aware training approach in FL-based IoMT networks.}
    \label{fig:Comparison_4_methods_fig_bin}
\end{figure}

\begin{figure}[!ht]
    \centering
    \includegraphics[width=\linewidth, height=4.3cm]{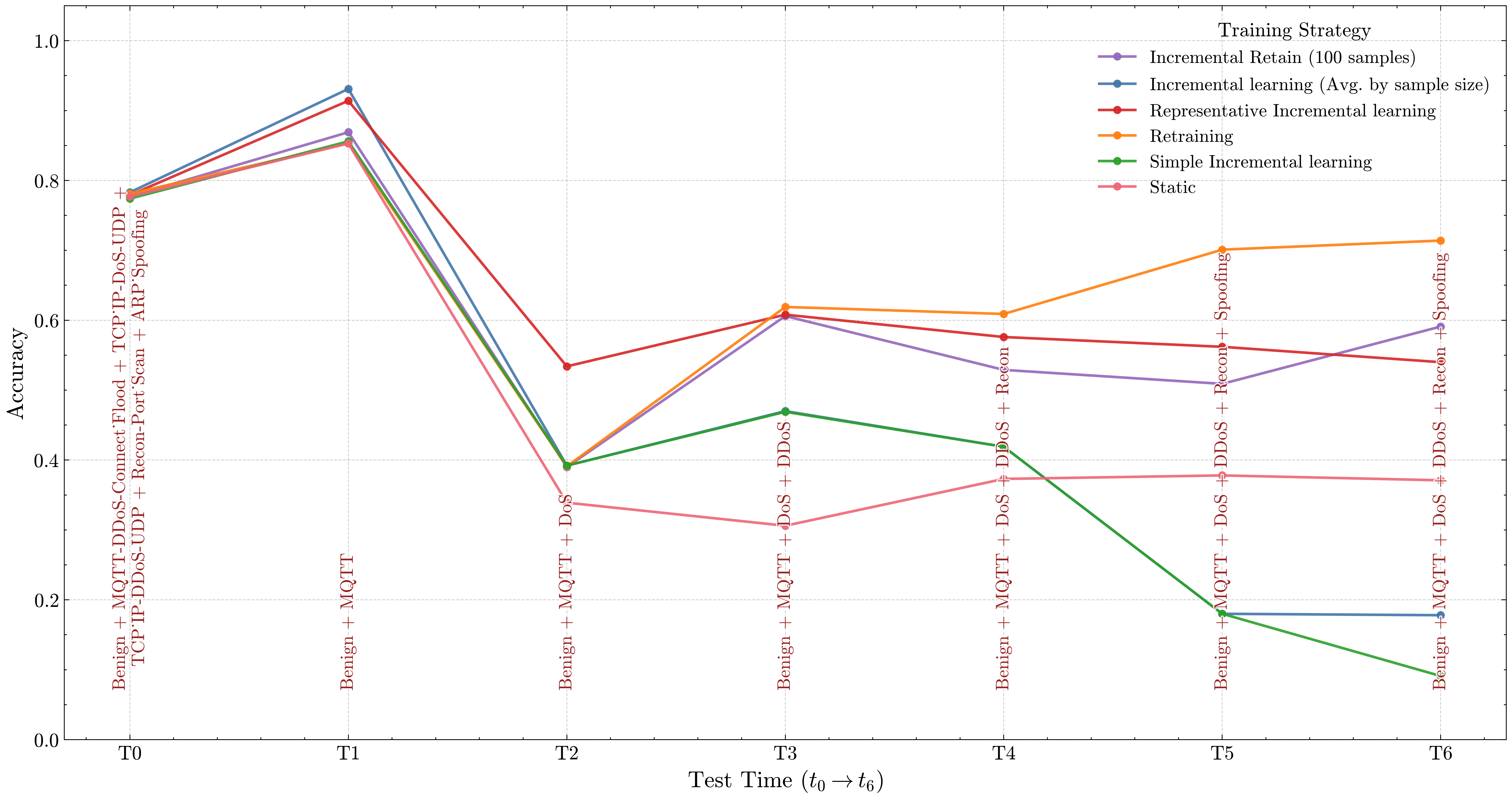}
    \caption{Six-class classification accuracy performance across timeline $t_{0}$–$t_{6}$ considering the main six training strategies}
    \label{fig:Comparison_5_methods_cat}
\end{figure}

Figure~\ref{fig:Comparison_4_methods_fig_bin} and \ref{fig:Comparison_5_methods_cat} visualizes the comparative performance across time periods, while Tables~\ref{tab:accuracy_bin_all},~\ref{tab:accuracy_cat_all},~\ref{tab:latency_bin_all}, and ~\ref{tab:latency_cat_all} summarise the quantitative findings.


\subsection{Accuracy under Concept Drift}
\begin{table}[t]
\centering
\caption{Binary classification accuracy (\%) across timeline $t_{1}$–$t_{6}$}
\label{tab:accuracy_bin_all}
\renewcommand{\arraystretch}{1.05}
\setlength{\tabcolsep}{3pt}
\scriptsize
\resizebox{\columnwidth}{!}{
\begin{tabular}{lccccccc}
\toprule
\textbf{Method} & $t_1$ & $t_2$ & $t_3$ & $t_4$ & $t_5$ & $t_6$ & \textbf{Avg.} \\
\midrule
Static & 97.85 & 68.90 & 61.73 & 64.75 & 63.69 & 63.78 & \textbf{70.12} \\
Cumulative Incremental Learning & 97.81 & 67.69 & 99.44 & 98.52 & 97.80 & 98.56 & \textbf{93.30} \\
Simple Incremental & 97.74 & 68.90 & 87.78 & 87.28 & 96.62 & 45.93 & \textbf{80.71} \\
Representative Incremental & 96.32 & 84.81 & 99.12 & 98.48 & 98.10 & 97.54 & \textbf{95.73} \\
Incremental Retain (100) & 97.76 & 68.62 & 95.16 & 98.14 & 95.01 & 96.81 & \textbf{91.92} \\
Incremental Retain (500) & 97.77 & 68.87 & 98.40 & 97.96 & 96.40 & 97.04 & \textbf{92.74} \\
Incremental Retain (1000) & 97.83 & 67.54 & 98.32 & 97.70 & 97.23 & 97.32 & \textbf{92.66} \\
Incremental Avg (EMA) & 97.83 & 68.85 & 86.14 & 81.29 & 97.01 & 58.71 & \textbf{81.64} \\
Incremental Avg (Samples) & 97.74 & 68.51 & 85.61 & 80.76 & 97.25 & 56.45 & \textbf{81.05} \\
Incremental Avg (Equal) & 97.83 & 68.89 & 85.88 & 81.46 & 96.99 & 58.96 & \textbf{81.67} \\
\bottomrule
\end{tabular}}
\end{table}

\begin{table}[t]
\centering
\scriptsize
\caption{Accuracy (\%) across timeline $t_0$--$t_6$ (6-class classification)}
\label{tab:accuracy_cat_all}
\resizebox{\columnwidth}{!}{%
\begin{tabular}{lcccccccc}
\toprule
\textbf{Method} & $t_0$ & $t_1$ & $t_2$ & $t_3$ & $t_4$ & $t_5$ & $t_6$ & \textbf{Avg.} \\
\midrule
Static & 77.7& 85.3& 33.9& 30.6& 37.3& 37.8 & 37.1 & \textbf{48.5} \\
Cumulative Incremental & 78.1 & 85.3 & 39.1 & 61.9 & 60.9 & 70.1 & 71.4 & \textbf{66.7} \\
Simple Incremental & 77.4 & 85.6 & 39.2 & 46.9 & 41.9 & 18.0 & 9.1 & \textbf{45.4} \\
Representative Incremental & 77.9 & 91.4 & 53.4 & 60.8 & 57.6 & 56.2 & 54.0 & \textbf{64.5} \\
Incremental Retain (100) & 77.7 & 86.9 & 39.0 & 60.6 & 52.9 & 50.9 & 59.1 & \textbf{61.0} \\
Incremental Retain (500) & 77.8 & 86.6 & 39.1 & 60.8 & 54.5 & 64.6 & 61.9 & \textbf{63.6} \\
Incremental Retain (1000) & 77.8 & 86.0 & 39.1 & 60.6 & 53.9 & 66.1 & 68.6 & \textbf{64.6} \\
Incremental learning (EMA) & 78.1 & 86.7 & 39.2 & 46.9 & 41.9 & 18.0 & 9.2 & \textbf{45.7} \\
Incremental learning (Sample) & 78.3 & 93.1 & 39.2 & 47.0 & 41.9 & 18.0 & 17.8 & \textbf{47.9} \\
Incremental learning (Equal) & 77.9 & 86.5 & 39.0 & 46.9 & 41.9 & 18.0 & 9.1 & \textbf{45.6} \\
\bottomrule
\end{tabular}%
}
\end{table}

Table~\ref{tab:accuracy_bin_all} and ~\ref{tab:accuracy_cat_all} presents the classification accuracy of all training strategies across timeline $t_{1}$–$t_{6}$ for binary and $t_{0}$–$t_{6}$ for six-class classification, respectively. In the binary setting, evaluation begins from $t_{1}$, the earliest point where both training and testing share the same attack families, whereas in the six-class setting, $t_{0}$ serves as the no-drift baseline. All subsequent time periods introduce new attack families, resulting in progressively stronger distributional shifts that allow us to assess each method’s robustness to sequential concept drift. Furthermore,($t_{5}$) and ($t_{6}$) contain the same attack families in both binary and 6-class settings because we aim to use ($t_{6}$) exclusively for evaluation since no training data is introduced at this stage. 

To ensure fair comparison across all learning strategies, the testing dataset at each time period \(t_j\) contains the \textit{same attack classes} for every method (static, cumulative, simple, representative, and retention based incremental learning). Although the underlying sample instances may differ due to random shuffling, the set of attack labels included in each \(t_j\) remains identical across all strategies. For example, the test split at \(t_2\) consistently includes the same Benign, DoS and MQTT attacks for every method, regardless of how each approach selects or retains training samples at \(t_1\) or \(t_2\). This ensures that performance differences arise solely from the training methodology rather than discrepancies in the evaluation data.


In binary classification (Table \ref{tab:accuracy_bin_all}), representative incremental learning achieves the highest average accuracy ($95.73\%$), slightly outperforming cumulative incremental learning ($93.30\%$) while requiring significantly less computational effort. retention-based approaches also perform strongly (91.92–92.74\%), indicating that preserving limited historical data is highly effective in stabilising binary decision boundaries under drift. In contrast, simple incremental learning and averaging-based incremental learning show very less average accuracy scores, i.e. (80.71–81.67\%) and experience large drops at later time periods, i.e. \(t_6\)(45.93-58.96\%),demonstrating that these strategies fail to preserve knowledge when attacks arrive sequentially (Figure~\ref{fig:Comparison_4_methods_fig_bin}).



For 6-class evaluation across $t_0$–$t_6$, cumulative incremental learning achieves the best average accuracy (66.7\%), followed closely by representative incremental learning (64.5\%) and retention by limited samples methods (63.6–64.6\%). These approaches effectively preserve knowledge of earlier attack families while adapting to new ones. Simple incremental learning again deteriorates rapidly after \(t_4\), falling to just 9\% at \(t_4\), highlighting its inability to retain multi-class decision boundaries. The Static model maintains good accuracy only at the no-drift baseline ($t_0$) but fails under sequential attack introduction, confirming its unsuitability for evolving IoMT environments. 

The results given in Figure \ref{fig:binary_mean_accruacy} approximate the similarity between the attack categories based on the binary classification results. It can be inferred that MQTT is a more distinct category compared to others. More specifically, its similarity to DoS and DDoS is low, as a model trained to detect MQTT performs worse on DoS and DDoS detection, with accuracy scores of 0.509 and 0.317, respectively. In Figure \ref{fig:Comparison_4_methods_fig_bin}, the binary non-stationary model starts with knowledge of the MQTT attack. However, the performance drops greatly at $t_2$ as the DoS category, a more dissimilar one to MQTT, is introduced into the testing data set. Once the model is trained (i.e., simple or cumulative incremental) with DoS at $t_2$, its performance at $t_3$ improves. Note that DDoS was introduced in this period. We consider that knowledge of DoS acquired in the previous period helps detect DDoS (i.e., the detection accuracy for DoS and DDoS pair is 0.996 in Figure \ref{fig:Comparison_4_methods_fig_bin}), thus improving detection performance at $t_3$. The similarity between DDoS and Recon can justify the performance stability at $t_4$. We conclude that the analysis given in Figure \ref{fig:binary_mean_accruacy} is instrumental in developing concept drift scenarios so that introducing more distinct classes causes performance drops.

\subsection{Latency and Efficiency Analysis}

Training and inference latencies, measured on the Organizational HPC node (\textit{AMD Threadripper 3970X, 128 GB RAM, NVIDIA RTX 3090 24GB}), 
are summarized in Table~\ref{tab:latency_bin_all} and ~\ref{tab:latency_cat_all}. We compare latency only for the primary training strategies and exclude the averaging-based incremental variants (equal, sample, EMA), since they use the same data flow and update procedure as simple incremental learning and therefore exhibit nearly identical latency characteristics. Training latency corresponds to the total wall-clock duration for model convergence, while inference latency measures the average prediction time in a time period.  

\begin{table}[t]
\centering
\caption{Per time period training, total training latency and total inference across timeline $t_{1}$–$t_{6}$ in binary classification in seconds}
\label{tab:latency_bin_all}
\renewcommand{\arraystretch}{1.05}
\setlength{\tabcolsep}{3pt}
\scriptsize
\resizebox{\columnwidth}{!}{
\begin{tabular}{lccccc|cc}
\toprule
\textbf{Method} & $t_1$ & $t_2$ & $t_3$ & $t_4$ & $t_5$ & \textbf{T. Training} & \textbf{Infer} \\
\midrule
Static & 94.67 & -- & -- & -- & -- & 94.67 & 2.07 \\
Cumulative Incremental Learn & 49.47 & 126.45 & 147.80 & 195.68 & 169.40 & 688.80 & 2.39 \\
Simple Incremental & 49.55 & 78.46 & 80.93 & 38.83 & 31.94 & 279.72 & 2.41 \\
Representative Incremental & 118.20 & 85.00 & 98.68 & 115.68 & 63.89 & 481.45 & 2.40 \\
Incremental Retain (100) & 51.33 & 69.43 & 82.20 & 29.87 & 22.98 & 255.80 & 2.49 \\ new values , 
Incremental Retain (500) & 50.92 & 71.26 & 121.82 & 35.64 & 32.03 & 311.67 & 2.52 \\
Incremental Retain (1000) & 52.30 & 118.20 & 83.58 & 48.38 & 37.79 & 340.24 & 2.52 \\
\bottomrule
\end{tabular}}
\end{table}

In binary classification Table~\ref{tab:latency_bin_all}, cumulative incremental learning incurs the largest computational cost, requiring 688.8 s in total—over 2.5× higher than Simple Incremental (279.72 s) and nearly 3× higher than incremental by 100 samples retention (255.8 s).
retention-based strategies (100–1000 samples) provide substantial time savings while preserving high accuracy, demonstrating that storing a small rehearsal buffer is computationally efficient yet effective.
Representative Incremental is more costly (481.45s) due to the inclusion of whole attack class representation from every category across timeline, but still considerably cheaper than cumulative incremental learning, which considers all the available classes.

\begin{table}[t]
\centering
\caption{ Per time period training latency , total training latency and total inference across timeline $t_{0}$–$t_{6}$ in seconds in 6-class classification.}
\label{tab:latency_cat_all}
\renewcommand{\arraystretch}{1.05}
\setlength{\tabcolsep}{3pt}
\scriptsize
\resizebox{\columnwidth}{!}{
\begin{tabular}{lcccccc|cc}
\toprule
\textbf{Method} & $t_0$ & $t_1$ & $t_2$ & $t_3$ & $t_4$ & $t_5$ & \textbf{T.Training} & \textbf{Infer} \\
\midrule
Static & 67.28 & -- & -- & -- & -- & -- & 67.28 & 2.03 \\
Cumulative Incremental & 72.45 & 79.83 & 82.13 & 88.04 & 93.26 & 87.83 & 603.54 & 2.40 \\
Simple Incremental & 70.62 & 73.80 & 78.42 & 83.76 & 80.21 & 75.00 & 311.81 & 2.27 \\
Representative Incremental & 68.31 & 73.40 & 79.20 & 84.53 & 87.60 & 79.68 & 422.72 & 2.30 \\
Incremental Retain (100) & 68.42 & 71.65 & 76.00 & 79.50 & 83.61 & 76.00 & 295.05 & 2.26 \\
Incremental Retain (500) & 68.80 & 72.33 & 77.30 & 80.60 & 85.33 & 78.00 & 302.28 & 2.23 \\
Incremental Retain (1000) & 69.50 & 73.10 & 78.20 & 82.00 & 86.90 & 78.79 & 318.49 & 2.27 \\
\bottomrule
\end{tabular}}
\end{table}

A similar trend is observed in 6-class classification Table~\ref{tab:latency_cat_all}.
Cumulative incremental learning accumulates the highest cost (603.5 s), while retention-based methods are significantly faster, ranging between 295–318 s. Simple incremental (311.8 s) offers the lowest latency among adaptive methods, but—consistent with the accuracy results, it suffers from severe drift at later time periods.
Representative incremental (422.7 s) again incurs a moderate additional cost but provides a strong accuracy–latency balance.

Across both binary and 6-class settings, the total inference latency remains nearly constant (2.0–2.4,s) for all training strategies.
This indicates that the runtime computational cost is dominated by the LSTM model architecture rather than the adaptation method itself, implying that all strategies are equally suitable for deployment on resource-rich IoMT aggregation nodes.


\subsection{Discussion}\label{sec:discussion}
In this study, we evaluated the performance of incremental learning strategies under a specific drift scenario for both binary and multi-class intrusion detection tasks. In the binary setting, the initial model is trained using only one attack type, MQTT, along with benign data. In contrast, the multi-class setting assumes that the model begins with one representative attack from each category. Although evaluations could be conducted under many alternative scenarios (e.g., varying the initial attack knowledge or altering the order in which attack types appear over time), we contemplate our results still offer valuable insights into how incremental learning can perform under drift in a federated learning environment for IoT networks. In future work, we plan to extend our experiments with additional scenarios and datasets.

In the multi-class experiments, we assumed that the initial model is trained with one attack type from every category. As a result, although new attack types are introduced in each period, the total number of classes remains constant over time. A more challenging and realistic non-stationary classification problem would involve introducing new classes in later periods in a federated learning setting \cite{dong2022federated}. Such classes may initially appear only on a small subset of clients and later become more widespread, reflecting evolving non-IID conditions. An effective intrusion detection system should be able to learn from these early but limited observations to mitigate their broader impact in subsequent stages. We plan to investigate class emergence and variation in non-stationary federated learning settings as part of future work.

In the current study, we assumed IID conditions across clients. While our experiments give initial insight about non-stationary model performance in FL settings, they do not fully reflect real-world environments, as the clients can observe varying attack types and categories over time, requiring more comprehensive analysis under varying non-IID situations.  Furthermore, intrusion detection datasets often contain minority classes, which is also the case in our dataset, and the effects of non-IID distributions can be more pronounced in these categories. We plan to explore the interactions between non-IID conditions and minority classes in our future work.

\section{Conclusions and Future Work}
\label{sec:conclusion}
In this work we investigated the robustness of federated learning (FL) models under sequential concept drift in IoMT networks using the CIC-IoMT2024 dataset.  Our initial attack to attack evaluation revealed that the MQTT and DDoS show the largest distributional divergence.We constructed a time periods based evaluation framework in which new attack classes or whole attack families (categories) were introduced gradually across timeline, enabling us to study the drift adaptation and efficiency trade-offs. We conducted experiments on both binary and 6-class classification tasks, which revealed that drift severity varies significantly across attack families.

Among all training strategies, cumulative incremental learning consistently achieved the highest accuracy across timeline in both binary and multiclass settings, confirming the effectiveness of full re-optimization when computational cost is not a constraint. However, several lightweight approaches, particularly representative incremental learning and retention based methods offered a strong balance between accuracy and efficiency, reducing training time by more than half while maintaining competitive performance.  Simple incremental learning showed vulnerability to accumulated drift, especially in later time periods, underscoring the importance of retaining or retaining historical knowledge.

Overall, the results highlight that FL strategies, particularly those incorporating selective class representation or samples retention offer a scalable and efficient pathway for defending IoT systems against evolving cyber threats, eliminating the need for frequent, costly retraining while preserving adaptability in non-stationary environments.


Future research shall explore adaptive drift detection mechanisms to trigger model updates dynamically instead of relying on fixed time periods. While the current study focuses on IID clients and a single deep learning architecture, future work will evaluate the proposed strategies in cross-device non-IID federated learning settings, consider alternative model architectures, and extend experiments to multimodal or additional real-world datasets to further improve robustness and deployment readiness.



\normalsize
\bibliography{references}


\end{document}